\documentclass[12pt]{iopart}

\newcommand{\sig}[1]{\sum^{\infty}_{#1=0}}

\newcommand{\negl}[1]{\left(1+{\rm{O}}\left( #1 \right)\right)}
\usepackage{amssymb}
\usepackage[dvips]{graphicx}

\begin{document}

\title[Method of Asymptotics beyond All Orders \\ and Restriction on Maps]
{Method of Asymptotics beyond All Orders \\ and Restriction on Maps}

\author{Shigeru Ajisaka
\footnote[3]{To
whom correspondence should be addressed (g00k0056@suou.waseda.jp)}
and Shuichi Tasaki 
}

\address{Advanced Institute for Complex Systems
and 
Department of Applied Physics,\\
School
of Science and Engineerings,
Waseda University,\\
3-4-1 Okubo, Shinjuku-ku,
Tokyo 169-8555,
Japan}

\begin{abstract}
The method of asymptotics beyond all orders (ABAO) is known to be a useful tool to investigate separatrix splitting of several maps. For a class of simplectic 
maps, the form of maps is shown to be restricted by the conditions for the ABAO method to work well. 
Moreover, we check that the standard map, the H\' enon map and the cubic map satisfy the restrcitions.
\\
\\
Mathmatics Subject Classification:
34C37,\ 37C29,\ 37E99,\ 40G10,\ 70K44
\end{abstract}



\maketitle
\section{Introduction}
The method of asymptotics beyond all orders (ABAO) is useful to investigate separatrix splitting of several maps such as the standard map\cite{Lazutkinf}, 
the H\'enon map\cite{Tovbis}, the cubic map\cite{NakamuraHamada} 
and the Harper map\cite{Ajisaka}.
In this paper, we examine sufficient conditions for the second-order 
simplectic difference equations to exhibit heteroclinic/homoclinic tangles,
to which the ABAO analysis is applicable.
We consider a simplectic discretization of the differential equation
$\frac{d^2}{dt^2}y(t)=f(y(t))$ with $f$ an entire function:
\begin{eqnarray}
y(t+\sigma)-y(t)&=&\sigma p(t)
\nonumber
\\
p(t+\sigma)-p(t)&=&\sigma f(y(t+\sigma))
\label{basic}
\end{eqnarray}
where $\sigma$ denotes a time step of discretization.
We shall show that, under certain conditions, the applicability of the ABAO 
analysis restricts the form of maps.

\section{ABAO analysis \label{ABAO}}
For sufficiently small $\sigma$, heteroclinic/homoclinic 
tangles of (\ref{basic}) are explained in terms of the
Stokes phenomenon and it can be studied by the ABAO analysis (see \cite{GLS94}
\cite{Gelfreich4} for more details).
The perturbative solution is valid only in a certain sector of
the complex time plane and the heteroclinic/homoclinic tangles are represented
by additional terms which appear when the solution is analytically continued to a different sector.
And the additional terms are picked up by the ABAO analysis.
The key idea of this method is to employ the so-called inner 
equation and to investigate it with the aid of 
the resurgence theory\cite{Ecall} and the Borel 
resummation. The inner equation magnifies the behavior of the solution 
near its singularities and bridges the solutions in different sectors.
The procedure is summarized as follows
\begin{itemize}
\item[1]:
Find a singularity $t_c$ of the perturbative solution in 
the complex time domain.
\item[2]:
Sum up the most divergent terms in each order solution.
The sum (called the inner solution) is valid in a sector, e.g., Re$(t)<{\rm Re}(t_c)$ 
and is given as the solution of the so-called inner equation.
\item[3]:
With the aid of the resurgence theory and the Borel resummation, the inner solution is 
analytically continued to the other sector\break Re$(t)>{\rm Re}(t_c)$ 
where it acquires new terms.
This completes the analytical continuation in a neighborhood of $t_c$.
\item[4]:
The solution which is valid far from $t_c$ is expanded as
a double expansion with respect to $\sigma$ and 
$\epsilon\sim e^{-a/\sigma} (a>0)$:
\begin{eqnarray}
y(t)=\sig{n} \sig{l} \sigma^{j_n} \epsilon^{n}y_{nl}(t)\sigma^l
\nonumber
\\
p(t)=\sig{n} \sig{l} \sigma^{j_n} \epsilon^{n}p_{nl}(t)\sigma^l
\label{expansion}
\end{eqnarray}
The quantities $\epsilon,j_n,y_{nl}(t),p_{nl}(t)$
are determined so that the asymptotic form of (\ref{expansion}) 
with respect to $z\equiv(t-t_c)/\sigma$ agrees with the solution of 
the inner equation obtained in the previous step.
\end{itemize}
As will be shown in Sec.4, the new terms added in the third step 
obey the linearized inner equation
(see \cite{Hakim} for more details).
On the other hand, if there appear heteroclinic/homoclinic tangles, 
the added terms should grow polynomially terms 
in the limit of $z\to\infty$.
We show that, under certain conditions, this property as well as the existence
of the inner equation restrict the maps.
Note that the inner equation was firstly used by Lazutkin\cite{Lazutkinf}
to derive the first crossing angle between the stable and unstable 
manifolds of the standard map, and by Kruskal and Segur\cite{Kruskal}, 
to study a singular perturbation problem of ordinary differential equations.

\section{Inner equation}
Let $t_c$ denote a singular point whose real part is smallest 
among all singularities.
In what follows, we discuss analytic continuation
from Re$(t_c)<0$, where the perturbative solution is assumed to be valid,
to Re$(t_c)>0$.
Let $(y_0(t,\sigma),p_0(t,\sigma))$ be the perturbative solution of (\ref{basic}):
\begin{eqnarray}
y_0(t,\sigma)=\sig{j}\sigma^j y_{0j}(t),\ \ p_0(t,\sigma)=
\sig{j}\sigma^j p_{0j}(t) \ .
\end{eqnarray}
For the maps studied by the ABAO analysis, it is observed that there exists a
non-negative integers $n$ and $m$ such that the order of a pole of $y_{0j}(t)$ 
($j\ge n$) at $t=t_c$ is $m+j-n$. In this case, the most divergent term
of $\sigma^j y_{0j}(\sigma z+t_c)$ for $z\to 0$ behaves like 
$\sigma^{n-m}/z^{m+j-n}$. Therefore, the following 
statement is expected to hold.
\medskip

\noindent
{\bf A:} The limit $\displaystyle \lim_{\sigma\to 0} \Phi_0(z,\sigma)=
\Phi_{00}(z)$ exists where 
$$
\Phi_0(z,\sigma)=
\Biggl(y_0(\sigma z+t_c)-\sum^{n-1}_{j=0} y_{0j}(\sigma z+t_c)\sigma^j\Biggr)
\sigma^{m-n}
$$

\medskip

In what follows, the condition A is assumed. Then, from the first equation 
of (1), the limit
\begin{eqnarray}
\Psi_{00}(z) &\equiv& \lim_{\sigma\to 0}\{\sigma^{r+1}p_0(z\sigma+t_c)-\sigma^r \Delta y_s(z\sigma+t_c)\}\cr
&=& \lim_{\sigma \to 0}\Delta \Phi_0(z,\sigma)=\Delta \Phi_{00}(z)
\end{eqnarray}
exists where $y_s(t)\equiv\sum^{n-1}_{j=0} y_{0j}(t)\sigma^j$, $r=m-n$ and 
$\Delta$ 
is the difference operator with respect to $z$: \hfil \break
\centerline{$\Delta g(z)\equiv g(z+1)-g(z)$.}

\noindent
By a similar argument, (4) and the second equation of (1) 
imply the existence of the limit:
\begin{eqnarray}
\lim_{\sigma \to 0}\left\{\sigma^{r+2} f
\left({\Phi_0(z+1,\sigma)\over \sigma^r}+y_s(\sigma(z+1)+t_c)\right)
-\sigma^r \Delta^2 y_s(\sigma z+t_c)\right\}
\nonumber \\
=\Delta \Psi_{00}(z)
\end{eqnarray}
Therefore, one may assume 

\medskip

\noindent
{\bf B:} The limit 
\begin{eqnarray}
S(a,z)\equiv\lim_{\sigma \to 0}\Biggl\{\sigma^{r+2} f\left({a\over \sigma^r}+y_s(\sigma(z+1)+t_c)\right)
-\sigma^r \Delta^2 y_s(\sigma z+t_c)\Biggr\}
\nonumber
\end{eqnarray}
exists and the convergence is uniform in $a$.

\medskip

\noindent
Then, from the above observation, one finally obtains the inner equation
\begin{eqnarray}
\Delta\Phi_{00}(z)&=&\Psi_{00}(z)
\nonumber
\\
\Delta\Psi_{00}(z)&=&S\left(\Phi_{00}(z+1),z\right)
\label{inner0}
\end{eqnarray}

An interesting observation follows from the above conditions.\\
\noindent{\bf{Proposition}}\\
If A and B$'$: $\exists\lim_{\sigma\to 0} \sigma^r y_s(z\sigma+t_c)\equiv 
\hat{y}_s(z)$ are satisfied and if $f(y)$ is a polynomial,
$f(y)$ should be quadratic or cubic.
\\ \\
\noindent{\it{Proof}}: \
Let $f(y)$ be the $N$th order polynomial, then
\begin{eqnarray*}
\sigma^{r+2}f\left(\sigma^{-r}\Phi_0(z+1,\sigma)+y_s(z+1,\sigma)\right)
\\
=\sigma^{r+2-Nr}
\left(\Phi_{00}(z+1)^N+\cdots
+\hat{y}_s(z+1)^N\right)
\times (1+\rm{O}(\sigma))
\end{eqnarray*}
It converges as $\sigma\to 0$, if and only if
$$(r+2)-rN=0$$
which admits only the following combinations.
$$(N,r)=(2,2),\ (3,1) \mskip 150 mu{\it Q.E.D.}$$

The first pair $(2,2)$ corresponds to the H\'enon map\cite{Tovbis} and
$(3,1)$ to the cubic map\cite{NakamuraHamada}.
Note that
the standard map is one of examples which non-trivially satisfy 
A and B with $r=0$\cite{Gelfreich99}.

\section{Linearized inner equation\label{linearized}}
As mentioned before, when the inner solutions are analytically continued to 
the other sector\break
Re$(t)>{\rm Re}(t_c)$, they acquires new terms.
Let $(\widetilde{\Phi},\widetilde{\Psi})$ be the analytical continuations of 
$(\Phi_{00},\Psi_{00})$, then they generally admit the expansions:
\begin{eqnarray}
\widetilde{\Phi}(z)=\sig{i}\Phi_{i0}(z) e(z)^i,\ 
\widetilde{\Psi}(z)=\sig{i}\Psi_{i0}(z) e(z)^i
\nonumber
\end{eqnarray}
where $e(z)=e^{-2\pi iz}$ for Im$(t_c) >0$ and
$e(z)=e^{2\pi iz}$ for Im$(t_c) <0$.
As $(\widetilde{\Phi}(z),\widetilde{\Psi}(z))$ again satisfy (\ref{inner0}),
$(\Phi_{10}(z),\Psi_{10}(z))$ obeys the linearized inner equation:
\begin{eqnarray}
\Delta \Phi_{10}(z)&=&\Psi_{10}(z)
\nonumber
\\
\Delta \Psi_{10}(z)&=&F\left(z \right)
 \Phi_{10}(z+1)
\label{linearized}\\
F(z)&\equiv& \left.{\partial \over \partial a}S(a,z)\right|_{a=\Phi_{00}(z+1)} \nonumber
\end{eqnarray}
When there exist heteroclinic/homoclinic tangles, 
$(\Phi_{10},\Psi_{10})$ would include finite polynomials of $z$.
Indeed, if they are expanded into a series like 
$\displaystyle{\sum_{n=1}^{\infty}\frac{c_n}{z^n}}$, 
the added terms vanish for $z\to\infty$ and the solutions of (\ref{basic}) 
would not acquire new terms in the other sector Re$(t)>{\rm Re}(t_c)$.
Note that the H\' enon map, the cubic map and the standard map have this 
property.
Therefore, $\Phi_{10}$ and $\Psi_{10}$ are expanded as
\begin{eqnarray}
\Phi_{10}(z) &=&
\sum_{i=-\infty}^{m_1}a_{i} z^i
,\ \ 
\Psi_{10}(z)=
\sum_{i=-\infty}^{n_1}b_{i} z^i
\nonumber
\end{eqnarray}
where $m_1>0,\ n_1>0,\ a_{m_1}\neq 0,\ a_{n_1}\neq 0$.
Then, the first equation of (\ref{linearized}) leads to
\begin{eqnarray}
1&=&\frac{\Delta\Phi_{10}(z)}{\Psi_{10}(z)}
\nonumber\\
&=&\frac{m_1a_{m_1}z^{m_1-1}\negl{\frac{1}{z}}}
{b_{n_1}z^{n_1}\negl{\frac{1}{z}}}
\nonumber\\
&=&m_1\frac{a_{m_1}}{b_{n_1}}z^{m_1-n_1-1}\negl{\frac{1}{z}}
\end{eqnarray}
and, thus,
\begin{eqnarray}
n_1-m_1=-1
\nonumber\\
\frac{b_{n_1}}{a_{m_1}}=m_1=n_1+1
\nonumber
\end{eqnarray}
On the other hand, from the second equation of (\ref{linearized}), we have
\begin{eqnarray}
F(z)&=&\frac{\Delta \Psi_{10}(z)}{\Phi_{10}(z+1)}
\nonumber
\\
&=&
\frac{n_1 b_{n_1}z^{n_1-1}\negl{\frac{1}{z}}}
{a_{m_1}z^{m_1}\negl{\frac{1}{z}}}
\nonumber
\\
&=&\frac{n_1 b_{n_1}}{a_{n_1}}z^{n_1-m_1-1} \negl{\frac{1}{z}}
\nonumber
\\
&=&\frac{n_1(n_1+1)}{z^2}\negl{\frac{1}{z}}
\end{eqnarray}
It is remarkable that the asymptotic expansion of $F(z)$ starts from $1/z^2$ and its coefficient is a product of successive natural numbers.
In short, if the solution of the linearized inner equation 
(\ref{linearized}) are sums of polynomials and inverse powers of $z$,
a further restriction is imposed on the map:
\begin{equation}
\lim_{z\to \infty} z^2 \left.{\partial \over \partial a}S(a,z)\right|_{a=\Phi_{00}(z+1)}=n_1(n_1+1)
\label{condition}
\end{equation}
where $n_1$ is a natural number.
The H\' enon, cubic and standard maps do satisfy (\ref{condition}),
respectively, with $n_1=3$, 2 and 1.
Note that the existence of the polynomial parts would be equivalent to the 
heteroclinic/homoclinic tangles.

\section{Conclusion}
We have briefly reviewed the ABAO analysis and shown that, 
for a class of simplectic maps, the form of maps is restricted 
by the existence of the inner equation, which is a main tool of the 
ABAO analysis, 
and asymptotic properties of the solutions of the linearized inner 
equation.
Namely, for simplectic maps (\ref{basic}) with an entire $f(y)$,
\begin{itemize}
\item
If $f(y)$ is a polynomial and the conditions A and B$'$ are satisfied, $f(y)$ should be
quadratic or cubic.
\item
Under the conditions A and B,
an additional restriction (\ref{condition}) should
be imposed for the map if the solution of the linearized inner equation
grows polynomially for $z\to \infty$
(that would imply the existence of heteroclinic/homoclinic tangles).
\end{itemize}
We would like to remark that the positivity of
the index $r$ studied in Sec.3 seems to be related to the existence of
the logarithmic branch point in the complex time domain.
Indeed, the H\' enon map and the cubic map (where $r>0$) 
do not have branch points and 
the standard map (where $r=0$) have a logarithmic branch point.
This aspect will be studied elsewhere.

\section*{Acknowledgments}

The authors thank Dr. T. Miyaguchi, Prof. K. Nakamura, Prof. M. Robnik, 
Prof. D. Sauzin and Prof. A. Shudo for 
fruitful discussions and useful comments.
This work is partially supported by a Grant-in-Aid for Scientific Research
of Priority Areas ``Control of Molecules in Intense Laser Fields'' and the
21st Century COE Program at Waseda University ``Holistic Research and Education
Center for Physics of Self-organization Systems'' both from Ministry of 
Education, Culture, Sports and Technology of Japan, by a Grant-in-Aid for 
Scientific Research (C) from the Japan Society of the Promotion of Science, 
as well as by Waseda University Grant for Special Research
Projects, Individual Research (2004A-161).


\begin{thebibliography}{xxxx}

\bibitem{Lazutkinf}
Lazutkin V. F., 
``Splitting of separatrices for the Chirikov's standard
map'' VINITI 6372/84  preprint (1984)[Russian]


\bibitem{Tovbis}
Tovbis A., Tsuchiya M. and Jaff\'e C.,
``Exponential asymptotic expansions and approximations of the unstable and stable manifolds of singularly 
perturbed systems with the H\'enon map as an example.''
{\sl  Chaos} {\bf{8}} 665-681 (1998);\ 
``Exponential asymptotic expansions and approximations of the
unstable and stable manifolds of the H\'enon map'', preprint (1994).


\bibitem{NakamuraHamada}
Nakamura K. and Hamada M.,
``Asymptotic expansion of homoclinic structures in a symplectic mapping.''
{\sl J. Phys. A} {\bf{29}} 7315-7327 (1996)


\bibitem{Ajisaka}
S.Ajisaka and S.Tasaki,
{\sl nlin.CD/0606030}


\bibitem{GLS94}
Lazutkin V. F., Gelfreich V., and Svanidze N.V.,
``A refined formula for the separatrix splitting for the standard map.''
{\sl Physica D} {\bf{71}(2)} 82-101 (1994)

\bibitem{Gelfreich4}
Gelfreich V. and Sauzin D.,
``Borel summation and splitting of separatrices for the H\'enon map.'' 
{\sl Annal. Instit. Fourier},{\bf{51}},513-567 (2001)
and references therein.


\bibitem{Ecall}
\'Ecalle J.,
{\it Les fonctions \'resurgentes},
{\bf{2}}
(Publ. Math. D'Orsay, Paris) (1981)


\bibitem{Hakim}
Hakim V. and Mallick K.,
``Exponentially small splitting of separatrices, matching in the complex plane and Borel summation.''
{\sl  Nonlinearity} {\bf{6}} 57-70 (1993)


\bibitem{Kruskal}
Kruskal M. D. and Segur H.,
``Asymptotics beyond all orders in a model of crystal growth.''
{\sl  Stud.Appl.Math} {\bf{85}} 129-181 (1991)


\bibitem{Gelfreich99}
Gelfreich V.,
``A proof of Exponentially Small Transversality
of the Separatrices for the Standard Map''
{\sl Comm. Math. Phys.} {\bf{201}(1)} 155-216 (1999)


\end{thebibliography}
\end{document}